\documentclass{ws-ijn}

\usepackage{multicol}
\usepackage[superscript]{cite}

\newcommand*{\half}{\frac{1}{2}}

\newcommand*{\avg}[1]{\langle #1 \rangle}
\newcommand*{\Avg}[1]{\biggl\langle #1 \biggr\rangle}

\begin{document}

\markboth{Patrick Snyder}
         {Modeling a nanocantilever based biosensor...}

\title{Modeling a nanocantilever based biosensor using a stochastically
       perturbed harmonic oscillator}

\author{Patrick Snyder and Amitabh Joshi}

\address{Department of Physics, Eastern Illinois University\\
         Charleston, Illinois 61920, USA\\
\email{joshi@eiu.edu}}

\author{Juan D. Serna}
\address{School of Mathematical and Natural Sciences,
              University of Arkansas at Monticello\\
              Monticello, Arkansas 71656, USA\\
\email{serna@uamont.edu}}

\maketitle


\begin{abstract}
Nanoscale biosensors are devices designed to detect analytes by combining
biological components and physicochemical detectors. One of the well known
methods of constructing these sensors is by using nanocantilevers. These
microscopic ‘diving boards’ are coated with binding probes that have an affinity
to a particular amino acid, enzyme or protein in living organisms. When these
probes attract some target particles, such as biomolecules, they change the
vibrating frequency of the cantilever. This process is random in nature and
produces fluctuations in the frequency and damping of the cantilever. In this
paper, we studied the effect of these fluctuations using a stochastically
perturbed classical harmonic oscillator.
\end{abstract}


\begin{multicols}{2}
\section{\label{sec:intro}Introduction}
Biosensors are composite devices consisting of biological sensing elements and
transducer systems. The working principle of these devices involves the binding
of the desired analyte to the biorecognition element fixed on a suitable support
matrix connected to a transducer. The binding of analytes causes changes in the
physical and/or chemical properties of the bioreceptive elements together with
the support matrix, which can then be sensed by a transducer to generate an
electrical signal. This generated signal quantifies the amount of analyte
deposited on the system. Classification of biosensors can be based on either
biorecognition mechanisms or the methodology of signal transduction\cite{Kumar}.

Nanobiosensors utilizing nanocantilevers can provide extreme sensitivity in the
detection of biomolecules (analytes) down to a single-particle
level\cite{Baselt,Ekinci}. Detecting particular biomolecules can help
researchers to recognize pathogens and diseases during clinical monitoring. Like
many other detectors, nanoscale biosensors are characterized by a quantity
called the \emph{dynamic range}, determined by the minimum mass detection limit
to the saturation limit in the detector\cite{Bellan,Ekinci}.

Nanocantilever biosensors are based on the mechanical motion of cantilevers. A
cantilever-based biosensor works as a tiny mechanical device whose mass
continually changes as biological analytes attach to it. The attachment of
analytes leads to a change in the resonant frequency of the device. The amount
of mass deposited on such detectors can be estimated by measuring the shift in
the resonance frequency of the resonator. These detectors possess high
sensitivity because of their intrinsic mass is small but the sticking of analyte
molecules causes significant mass change and hence produces a very large change
in the resonant frequency (typically in the range of 15 to 20 \%). The quality
factor of such nanodevices is also large and that further adds to their
sensitivity of detection\cite{Bellan,Fritz}.

Recently, a slightly different kind of nanocantilever has emerged. This kind of
nanocantilever-based biosensor device contains nanofluidic channels that are
used to detect small mass species, e.g., cells, proteins, etc. The nanofluidic
channels allow one to operate under a low pressure background which enhances the
quality factor of the detector\cite{Burg}. In order to measure the change in
frequency of nanocantilevers, it is required that the system parameters remain
unaltered. Ideally speaking, once the analyte is attached to the nanocantilever,
it should not detach or move from it\cite{Atalaya}. Such an
attachment-detachment or adsorption-desorption of the analyte
particles\cite{Cleland}, including their random striking on the
detector\cite{Atalaya}, which involves momentum transfer to the
nanocantilevers\cite{Ekinci}, are the main factors for the spread or broadening
of the resonant frequency along with shifts in the frequency.

In order to develop a theoretical model related to the detection process for
such generic nanocantilevers (a few examples are shown in Fig.~\ref{Fig:01}), it
is important to keep all the macroscopic and microscopic factors involved in the
detection process in mind\cite{Bellan}. For such a tiny system having dimensions
of several tens to hundred nanometers, it is indeed possible to analyze some
variables macroscopically, but the proper recognition of such variables is quite
critical, which is less obvious and would be negligible if one is not working on
the nanoscale. Because amino acids are one of the main bioreceptors used in the
nanocantilever-based biosensors, the affinity and specificity of amino acid
sequences (peptides, proteins, etc) are a very important factor in such
analysis\cite{Bellan}. These amino acids are critically dependent on ligands to
which they are attached and responsible for the determination of the geometric
dimensions of amino acid sequences. A change in the geometry could cause
receptor amino acid sequences to become inactive or even initiate other events
like the nonlinear expansion or the stochastic motion of the cantilever. These
irregular events (noise) need to be taken into account when nanocantilever
biosensors are modeled mathematically.

This paper is organized as follows: In section~\ref{sec:massDeposition}, we give
a brief description of how mass deposition causes frequency shifts in the
harmonic oscillator. In section~\ref{sec:sModels}, we model nanocantilevers
mathematically using a stochastically perturbed harmonic oscillator. We explain
how noise introduces measurable random frequency fluctuations in the oscillating
nanocantilever that are proportional to the mass deposited on its surface. This
feature is used by biosensors as the detecting mechanism\cite{Kumar, Bellan}.
Finally, some concluding remarks are presented in section~\ref{sec:summary}.
\begin{figurehere}
  \centerline{\epsfig{file=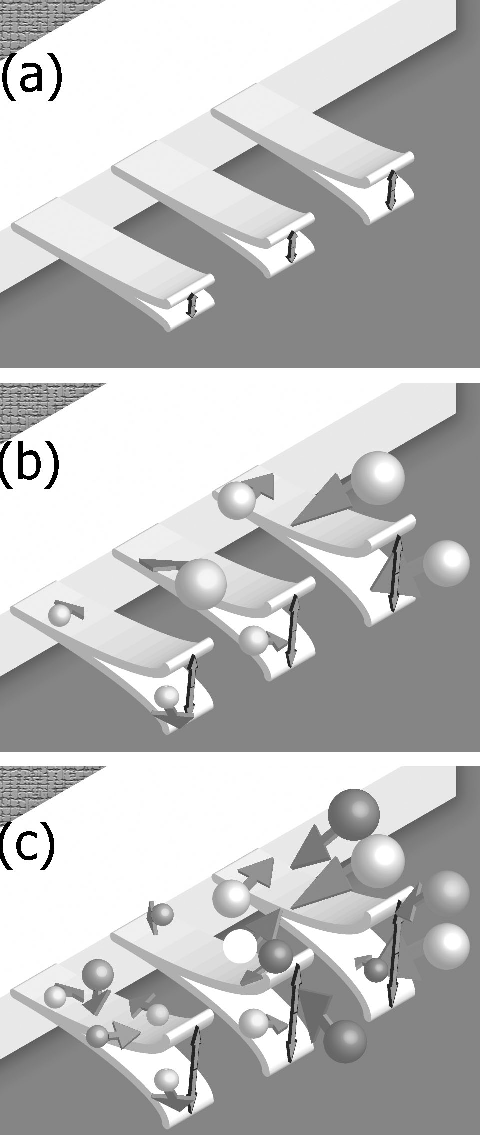,width=3.8cm}}
  \caption{Nanocantilever based biosensors in (a) vacuum, (b) gas and
  (c) solution environments.}
  \label{Fig:01}
\end{figurehere}

\section{\label{sec:massDeposition}Mass {D}eposition and {F}requency {S}hift}

Nanocantilevers can be simply modeled as damped, driven harmonic oscillators.
The oscillating mechanism we consider in this work consists of a mass-spring
system, and its equation of motion will be used to describe the
\emph{deterministic} behavior of nanocantilevers. As it is well known, this
equation includes damping, intrinsic frequency and  driving terms. Random
damping and random frequency terms due to fluctuation mechanisms in the
oscillator change its intrinsic resonance frequency. This is further discussed
in sections~\ref{ssec:damped} and~\ref{ssec:frequency}. If these frequency
changes can be measured, then the oscillators (nanocantilevers) can be used to
detect simple physical phenomena such as mass deposition on them.

For a nanocantilever, it is common to measure changes in its frequency of
resonance. This frequency shift is directly proportional to the mass deposited
on it as we discuss in the following. For a mass-spring system with mass $m$ and
spring constant $k$, the characteristic angular frequency under simple harmonic
motion is given by $\omega_0 = \sqrt{k/m}$. A small change in the mass of the
system due to mass deposition modifies the frequency of oscillation of the
system. Taking finite differences of the angular frequency $\Delta\omega_0$ with
respect to mass and solving for $\Delta m$, we get
\begin{equation}\label{Eq:massDep01}
  \Delta m = -\frac{2 m}{\omega_0} \Delta\omega_0.
\end{equation}
From this equation it is clear that the deposited mass $\Delta m$ can be
estimated by measuring $\Delta\omega_0$. For this equation to make physical
sense, we assume that the elastic and geometric properties of the nanocantilever
remain unchanged after a  small mass deposition.

The resonant frequency measurement of the nanocantilever can be done with a
microscopic scanning laser Doppler vibrometer setup\cite{Gupta,Lee}. The setup
involves a collimated laser beam passing through a polarizing beam splitter and
focused onto a cantilever with the help of a microscope objective. The reflected
beam passes through the beam splitter to fall on a photo detector. The typical
spot size of the laser beam is about 1--2$\,\mu\mbox{m}$. The frequency response
of the nanocantilever can be measured using a fixed amplitude function generator
along with a lock-in amplifier signal detection technique. The signal from the
function generator (the external driving force) on the nanocantilever can be
applied with the help of an additional electrode situated near to it and keeping
the nanocantilever grounded so that electrostatic actuation can be achieved.

When a nanocantilever with physical dimensions of length $\sim 5\,\mu\mbox{m}$,
width $\sim 1.5\,\mu\mbox{m}$, and thickness $\sim 30\,\mbox{nm}$ is used, the
resonant frequency of the nanocantilever falls in the range 1--2$\,\mbox{MHz}$.
Also, the spring constant of the nanocantilever is 0.005--0.010$\,\mbox{N/m}$.
If the analyte is a single vaccinia virus particle, its mass is about 5--8
femtograms. The concentration of such virus particles in an aqueous solution is
$\sim 10^9\,\mbox{PFU/ml}$ , where PFU means plaque forming in the virus sample.
The mass of DNA in a single cell is of the order of 6--7 picograms and the size
(Plasmid DNA) is about 10--20 nanometers.

When analyte molecules are deposited on a nanocantilever, there is a momentum
exchange between the oscillator and the particles colliding with it. This is
called a process of random collisions\cite{Ekinci}. These collisions bring a
random dragging force that affects the nanocantilever motion. If the
concentration of analyte molecules is low, then this dragging force can be
attributed to a molecular drag proportional to the molecular velocity and
contributing to the dissipative part of the equation of motion of the
oscillator. This dragging force can be written as $F_{\mathrm{d}} = b(dx/dt)$
and corresponds to the damping term of the oscillator's equation (e.g.,
Eq.~(\ref{Eq:DHO}) of section~\ref{sec:sModels}, in which $b$ is the damping
constant and $dx/dt$ represents the speed of the particle). When the
concentration of analyte molecules is low, we can say that $F_{\mathrm{d}}$ is
proportional to the number of analyte particles $N_{\mathrm{a}}$ striking the
nanocantilever per second. However, if the dragging force is random, it can be
described by the term $\xi(t)(dx/dt)$, in which the parameter $\xi(t)$
characterizes the randomness. The term brings random damping to the
nanocantilever oscillator\cite{Ekinci}. This analysis is applicable  for a
generic nanocantilever system. Another possible type of dragging is called
inertial dragging, and it is proportional to the molecular acceleration. We will
not further discuss this case here.

Other phenomenon taking place in this system is the adsorption-desorption of the
analyte molecules on the nanocantilever surface\cite{Cleland,Ekinci}. Such
thermally driven effect produces random frequency fluctuations in the
oscillator\cite{Cleland,Ekinci,Dykman}. The process of adsorption-desorption can
be modeled by a molecular flux-dependent adsorption rate and a thermally
activated rate of desorption\cite{Cleland,Ekinci}. Another interesting mechanism
that produces frequency fluctuations is the one occurring with nanocantilevers
having nanofluidic channels. In that case, the fluctuations are produced by the
diffusion of adsorbed particles along the nanofluidic channel inside the
vibrating nanocantilever. As the particle diffuses, the resonance frequency
changes according to the relative amplitude of the vibrating mode at the
location of the particle\cite{Atalaya}. All of these fluctuations or random
stochastic processes mentioned above can be incorporated as delta-correlated or
exponentially correlated noise processes  in the modeling equations.

\section{\label{sec:sModels}Stochastically {P}erturbed {H}armonic {O}scillator}

Let us consider the simple case of a nanocantilever described by a damped
harmonic oscillator of mass $m$ and spring constant $k$, driven by a sinusoidal
external force. The differential equation describing the motion is
\begin{equation}\label{Eq:DHO}
  m \frac{d^2 x}{d t^2} + b \frac{dx}{dt} + k x = F_0 \sin(\omega t),
\end{equation}
where $F_0$ and $\omega$ are the amplitude and frequency of the external driving
force, respectively, and $b$ is a positive damping coefficient.

If a system is subjected to both random and periodic forces, the well-known
phenomenon of stochastic resonance (SR) may emerge\cite{Gammaitoni,Gitterman}.
In the following, we will describe the effects of stochastic fluctuations in the
damping and frequency terms for the equation governing the dynamics of the
nanocantilever. These effects can be incorporated in~(\ref{Eq:DHO}) as
\emph{multiplicative} noise $\xi(t)$, such that the new equations of motion can
be written as the stochastic differential equations (SDE)
\begin{equation}\label{Eq:RDHO}
  \frac{d^2 x}{d t^2} + 2\beta[1 + \xi(t)]\frac{dx}{dt}
                      + \omega_0^2 x = A \sin(\omega t)
\end{equation}
for the random damping, and
\begin{equation}\label{Eq:RFHO}
  \frac{d^2 x}{d t^2} + 2\beta\frac{dx}{dt}
                      + \omega_0^2 [1 + \xi(t)] x = A \sin(\omega t)
\end{equation}
for the random frequency. Here $\beta \equiv b/2m$ is the damping parameter,
$\omega_0 = \sqrt{k/m}$ is the characteristic angular frequency in the absence
of damping, and $A = F_0/m$.

As discussed above, the nanocantilever can be used to estimate the weight of
individual nanoparticles, viruses, or protein molecules in an aqueous medium.
However, there is a fast dissipation of the nanocantilever energy due to viscous
dragging. If this aqueous medium is placed inside the cantilever using
nanofluidic channels, then the viscous dragging could be almost eliminated
\cite{Lee}. As an example, the molar concentration of water is
$55.5\,\mbox{mol/L}$ but the molar concentration of proteins in an E. Coli
bacteria is about $100\,\mbox{nmol/L}$. Protein size varies from $1\,\mbox{nm}$
to $5\,\mbox{nm}$, and the mass from $5,000\,\mbox{amu}$ to
$500,000\,\mbox{amu}$. The typical molecular separation goes from
$1.18\,\mbox{nm}$ (1 M solution) to $1.18\,\mu\mbox{m}$ (1 nM
solution)\cite{Erickson}. In this situation, both water molecules and analyte
particles $N_{\mathrm{a}}$ are randomly striking at the nanocantilever. At
thermal equilibrium, the translational kinetic energy of water molecules is
comparable to that of analytes (larger molecules). Since the molar concentration
of water is larger, then the random fluctuations due to particles striking on
the cantilever are dominated by water molecules, which eventually produce the
damping.

The random variable $\xi(t)$ will be considered as both Gaussian white noise
with the correlator
\begin{equation}\label{Eq:wNoise}
  \avg{\xi(t)\xi(t')} = D \delta(t-t'),
\end{equation}
and colored noise with exponential correlator
\begin{equation}\label{Eq:cNoise}
  \avg{\xi(t)\xi(t')} = \alpha^2\,e^{-\lambda |t - t'|}.
\end{equation}
The parameters $D$ and $\alpha$ represent the white and colored noise
strengths, respectively, and $\lambda$ the correlational decay rate. A special
case of the colored noise is the symmetrical dichotomous noise (random
telegraph signal) for which $\xi(t)$ can take one of the values $\xi = \pm
\alpha$, and the average waiting time for each of these states is
$\lambda^{-1}$. Under the limiting conditions: $\alpha^{2}\rightarrow \infty$
and $\lambda \rightarrow \infty$, Eq.~(\ref{Eq:cNoise}) reduces
to~(\ref{Eq:wNoise}), provided $\alpha^2 / \lambda = D$ (see
Ref.~\refcite{Gitterman}).

\subsection{\label{ssec:damped}Nanocantilever with random damping}

In this subsection we study the effect of random fluctuations produced by
particles striking on the cantilever in terms of the damping produced by them.
We call this phenomenon random damping. In the absence of an external driving
force ($A=0$) the equation of motion~(\ref{Eq:RDHO}) takes the form
\begin{equation}\label{Eq:OS1}
  \hat{O}_{\mathrm{D}}\,\{x\} = - 2\,\beta\,\xi(t) \frac{dx}{dt},
\end{equation}
where the operator $\hat{O}_{\mathrm{D}}$ is defined by the following expression
\begin{equation}\label{Eq:OS2}
  \hat{O}_{\mathrm{D}} \equiv \frac{d^2}{d t^2} + 2\beta \frac{d}{dt} +
\omega_0^2.
\end{equation}

In multiplicative stochastic processes involving terms like $\xi(t)x(t)$, the
selection of $t$ to find the appropriate $x(t)$ depends on the characteristic
noise correlation time for the variable $\xi(t)$. This gives rise to the
Ito-Strantonovich dilemma. If we take the asymptotic limit for the correlation
time, the solutions of the SDEs~(\ref{Eq:RDHO}) and~(\ref{Eq:RFHO}) are of the
Strantonovich form. On the other hand, to get solutions of the Ito form, it is
mandatory to have a correlation time of exactly zero. Also, it should be noted
that for any external noise source in the system, the Strantonovich formulation
works well\cite{Kampen}.

To get solutions of the Strantonovich form, Eq.~(\ref{Eq:OS2}) is recasted into
an integro-differential equation using the technique described in
Ref.~\refcite{Gitterman}. For this purpose, we apply the inverse operator
$O_{\mathrm{D}}^{-1}$ to~(\ref{Eq:OS1}) to get
\begin{equation}\label{Eq:OI1}
  x = - \hat{O}_{\mathrm{D}}^{-1} \left\{2\,\beta\,\xi(t) \frac{dx}{dt}\right\},
\end{equation}
The inverse operator $\hat{O}_{\mathrm{D}}^{-1}$ is an integral operator and can
be determined using~(\ref{Eq:OS2}), so that
\begin{eqnarray}\label{Eq:OI2}
  \hat{O}_{\mathrm{D}}^{-1}\,\{g\} \equiv \hspace{6cm} && \nonumber \\
  \frac {1}{\omega_0'}\int_0^t e^{-\beta(t-t')}
  \sin[\omega_0'(t-t')]g(t')dt', \hspace{1cm}&&
\end{eqnarray}
with $\omega_0' = \sqrt {\omega ^{2}-\beta^2}$. If we identify the function $g$
as the one given inside the curly braces in~(\ref{Eq:OI1}), then it is not
difficult to obtain
\begin{eqnarray}\label{Eq:x01}
  x(t) = -\frac {2\beta}{\omega_0'}\int_{0}^{t}
  \bigg\{ e^{-\beta(t-t')} \hspace{3cm} && \nonumber \\
  \sin[\omega_0'(t-t')]\xi(t')\frac{dx(t')}{dt'} \bigg\} dt', \hspace{1cm}&&
\end{eqnarray}
and
\begin{eqnarray}\label{Eq:x02}
  \frac{d}{dt}x(t) = \frac {2\beta}{\omega_0'}\int_{0}^{t}
  e^{-\beta(t-t')}\xi(t') \frac{dx(t')}{dt'} \hspace{2cm} && \nonumber \\
  \big\{\beta \sin[\omega_0'(t-t')]- \omega_0'\cos[\omega_0'(t-t')]\big\} dt'.
  \hspace{0.5cm} &&
\end{eqnarray}
Now, upon substituting~(\ref{Eq:OS1}) and~(\ref{Eq:OS2}) into~(\ref{Eq:x01})
and~(\ref{Eq:x02}), we may get
\begin{eqnarray}\label{Eq:x03}
\bigg\{\frac{d^2}{d t^2} + 2\beta \frac{d}{dt}
  + \omega_0^2 \bigg\}x(t) = \hspace{3.5cm} && \nonumber \\
  -\frac{4\beta^2}{\omega_0'}\int_{0}^{t}
  e^{-\beta(t-t')}\xi(t)\xi(t')\frac{dx(t')}{dt'} \hspace{2cm} && \nonumber \\
  \{ \beta \sin[\omega_0'(t-t')]-\omega_0' \cos[\omega_0'(t-t')] \} dt'.
  \hspace{0.5cm} &&
\end{eqnarray}

If the random damping is produced by the delta-correlated white
noise~(\ref{Eq:wNoise}), then~(\ref{Eq:x03}) reduces to
\begin{equation}\label{Eq:AvgDRHO}
\frac{d^2}{d t^2}\avg{x} + 2\beta (1 - 2\beta D)\frac{d}{dt}\avg{x}
                         + \omega_0^2 \avg{x} = 0.
\end{equation}
Here, we have use the fact that averages split in the form\cite{Bourret}
\begin{equation}\label{Eq:splitAvg}
  \Avg{\xi(t)\xi(t')\frac{dx(t')}{dt'}}
           = \avg{\xi(t)\xi(t')} \Avg{\frac{dx(t')}{dt'}}.
\end{equation}

Equation~(\ref{Eq:AvgDRHO}) models the \emph{collisional} damping produced by
analytes (in an aqueous medium, which could be within the cantilever in a
micro/nanofludic channel) getting stuck on the nanocantilever. If $2\beta D <
1$, the presence of white noise will lead to a damping decrease (weak noise). On
the other hand, if $2\beta D > 1$, the effective damping turns negative
increasing the amplitude of oscillation $\avg{x}$ and leading to instabilities
in the dynamics of the nanocantilever (strong noise)\cite{Gitterman}.

When the nanocantilever is driven by a sinusoidal external force $A\sin(\omega
t)$ (see Eq.~(\ref{Eq:RDHO})), a solution to~(\ref{Eq:AvgDRHO}) can be written
as
\begin{equation}\label{Eq:AvgDRHOsol}
  \avg{x} = B \sin(\omega t + \varphi).
\end{equation}
Upon substituting~(\ref{Eq:AvgDRHOsol}) into~(\ref{Eq:AvgDRHO}), we can solve
for the amplitude $B$ and get the following expression
\begin{equation}\label{Eq:B}
  B = \frac{A}
      {\left[(\omega^2 - \omega_0^2)^2 +
      4\beta^2\omega^2(1 - 2\beta D)^2\right]^{1/2}}.
\end{equation}

The presence of noise in the system brings fluctuations in the nanocantilever
frequency, resulting in spectral broadening. Hence it becomes necessary to
determine the minimum measurable frequency shift that can be observed in this
noisy environment. The spread in the frequency $\delta\omega_0$ can be obtained
by integrating the spectral density of fluctuations $S(\omega)$
\begin{equation}\label{Eq:SDF1}
  \delta\omega_0 \approx
  \left[\int_{\omega_0 - \Delta\omega_0}^{\omega_0 + \Delta\omega_0}
  S(\omega) d\omega \right]^{1/2}.
\end{equation}
Here, we assume that a measurement of the nanocantilever frequency was done with
a square-shaped transfer function over the bandwidth $2\Delta\omega_0$, and
centered at $\omega_0$. Equation~(\ref{Eq:SDF1}) is an estimate for any real
system. The spectral density $S(\omega)$ is determined by the nature of the
noise present in the system, and can be determined by taking the Fourier
transform of the white noise correlator\cite{Ekinci}
\begin{equation}\label{Eq:SDF}
  S(\omega) = \int_{-\infty}^{\infty} \avg{\xi(t)\xi(0)}
  \,e^{-i\omega t} dt.
\end{equation}

At equilibrium and under resonance, the second moment of the nanocantilever
displacement, $\avg{x^2}$, satisfies the relation\cite{Ekinci}
\begin{equation}\label{Eq:2Moment}
  \half m \omega_0^2 \avg{x^2} = \half \kappa T,
\end{equation}
where $\kappa$ is the Boltzmann constant and $T$ is the absolute temperature at
resonance. By using~(\ref{Eq:SDF}), we can find the spectral density
corresponding to these displacement fluctuations
\begin{equation}\label{Eq:Sx}
  S_{\avg{x^2}}= \frac{\mbox{const.} \times S(\omega)}
        {\left[(\omega^2 - \omega_0^2)^2 + 4\beta^2\omega^2
        (1 - 2\beta D)^2\right]^{1/2}}.
\end{equation}

There are two damping mechanisms present in~(\ref{Eq:Sx}): The thermo-mechanical
fluctuations governed by the damping parameter $\beta$, and the momentum
exchanged with white noise, whose spectral density is proportional to $\beta D$.
In particular, that momentum exchange is responsible for taking the system into
resonant states. The plot in Fig.~\ref{Fig:01} is for $S(x^2)/S(\omega)$ vs.
$\omega $ and five curves are drawn for different values of $\beta$. The
peculiarity in these resonance curves is that at resonance ($\omega=\omega_0$),
the peak height increases with increasing noise strength $D$, which is
counterintuitive. The peaks are appearing at the resonance condition
($\omega=\omega_0$) but their heights are dependent on the value $D$.
Alternatively, if we keep ($\omega-\omega_0$) as a fixed quantity
in~(\ref{Eq:Sx}) and plot the expression with respect to $D$, we get resonance
peaks located at $D=(2\beta)^{-1}$. This behavior resembles the stochastic
resonance phenomenon\cite{Joshi} in which the height of a peak (intensity)
goes up as noise strength increases up to a certain value and then goes down, a
counterintuitive observation (see Fig.~\ref{Fig:02}).

\begin{figurehere}
  \centerline{\epsfig{file=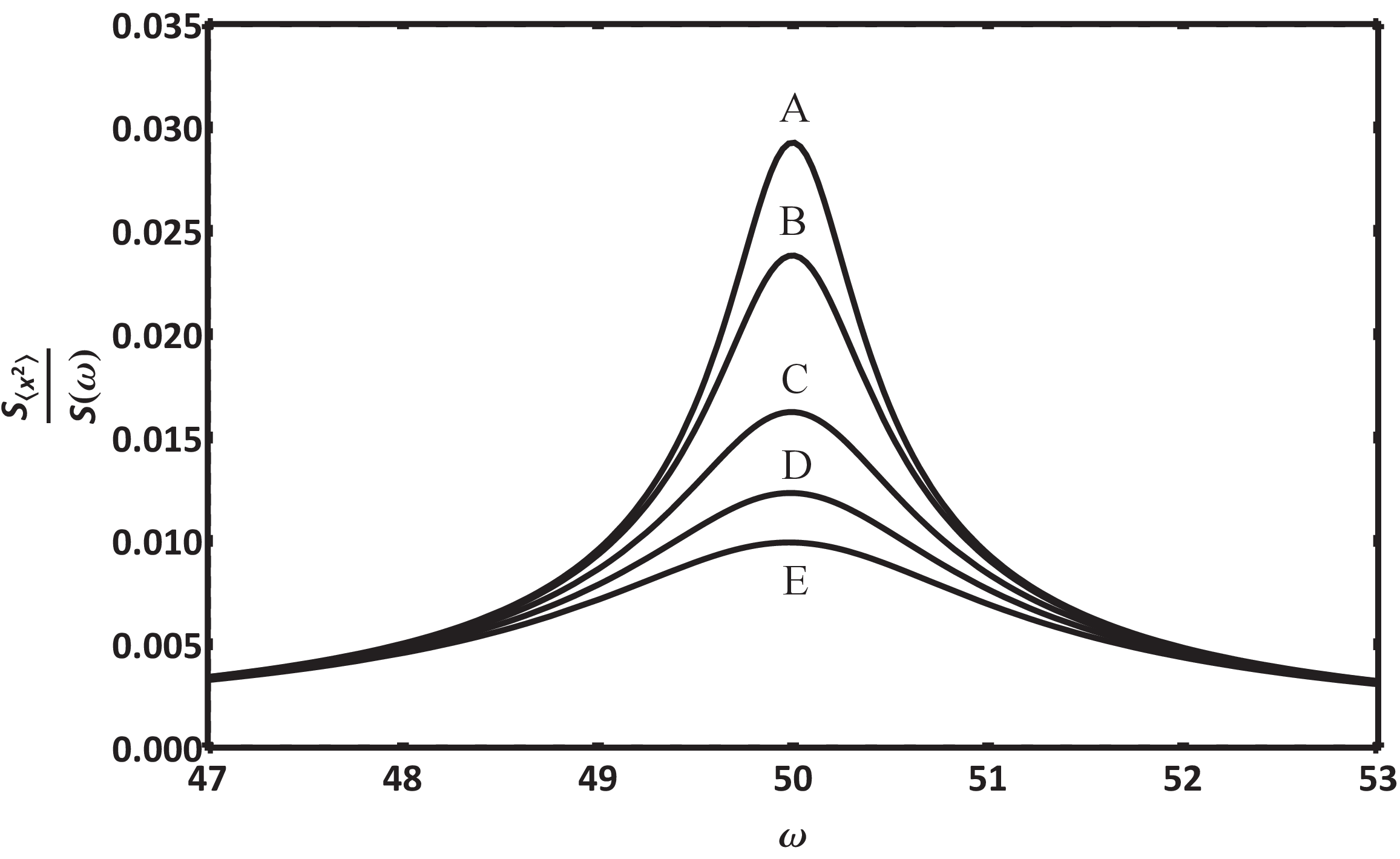,width=8.3cm}}
  \caption{Plot of $S_{\avg{x^2}}/S(\omega)$ as a function of the
  driving field frequency $\omega$, with parameter conditions: $\omega_0=50$
  and $\beta=1.4$ (see Eq.~(\ref{Eq:Sx})). Curves A, B, C, D, and E are for
  the diffusion parameter $D = 0.27,\,0.25,\,0.20,\,0.15$ and $0.10$,
  respectively.}
  \label{Fig:02}
\end{figurehere}

\noindent This counterintuitivly phenomenon of stochastic resonance is present
due to the fact that the ``unwanted'' noise coherently adds up to the external
force signal, increasing the signal to noise ratio instead of reducing
it\cite{Gammaitoni}. From~(\ref{Fig:01}), it is clear that, for these
nanocantilever based biosensors, the resonance peak is not sharp but has a
finite width. The peak width is determined by the intrinsic damping of the
nanocantilever, $\beta$, and the stochastic fluctuations caused by analyte
molecules sticking on it. To measure the precise shift in the resonance
frequency, the sharpest possible peak from the figure should be the appropriate
choice. Therefore, knowing the values of $\beta$ and $D$, and estimating the
mass deposition with accuracy become critical in designing suitable
nanocantilever based biosensors.

The response curve obtained from~(\ref{Eq:Sx}) and depicted in Fig.~\ref{Fig:02}
can be reproduced experimentally by measuring the frequency response curve with
and without the presence of fluctuations (introduced by analytes in an aqueous
medium). A brief discussion about such an experiment is given in
section~\ref{sec:massDeposition}. The damping parameter $\beta$ of the
nanocantilever can be measured by observing the decay of oscillations of the
nanocantilever after it has been excited by the driving force and then
disconnected. The value of $D$ can also be figured out at some known temperature
$T$ using widths of measured frequency curves (like in Fig.~\ref{Fig:02}) and
the $\beta$ value.

\subsection{\label{ssec:frequency}Nanocantilever with random fluctuations in
frequency}

In the absence of an external driving force ($A=0$) equation~(\ref{Eq:RFHO})
takes the form
\begin{equation}\label{Eq:RFHO2}
  \frac{d^2 x}{d t^2} + 2\beta\frac{dx}{dt}
                      + \omega_0^2 [1 + \xi(t)] x = 0.
\end{equation}
Here, the multiplicative noise $\xi(t)$ introduces random fluctuations in the
nanocantilever frequency. It can be shown that, if the noise is white noise with
correlator~(\ref{Eq:wNoise}), the first moment of the oscillator is not affected
by the noise and $\avg{x(t)}= x(t)$ (see Ref.~\refcite{Gitterman}). On the other
hand, for the exponentially correlated noise~(\ref{Eq:cNoise}), the first moment
$\avg{x}$ equation can be obtained by defining
\begin{equation}\label{Eq:OE1}
  \hat{O}_{\mathrm{E}}\,\{x\} = - \omega_0^2 \,\xi(t) x,
\end{equation}
where the operator $\hat{O}_{\mathrm{E}}$ stands for
\begin{equation}\label{Eq:OE2}
  \hat{O}_{\mathrm{E}} \equiv \frac{d^2}{d t^2} + \omega_0^2.
\end{equation}
By applying the inverse operator $O_{\mathrm{E}}^{-1}$ to~(\ref{Eq:RFHO2}) we
obtain
\begin{equation}\label{Eq:xe01}
  x = - \hat{O}_{\mathrm{E}}^{-1} \{ \omega_0^2\,\xi(t) dx \},
\end{equation}
Again, $O_{\mathrm{E}}^{-1}$ is an integral operator that can be determined
using~(\ref{Eq:OE2})
\begin{equation}\label{Eq:OEI02}
  \hat{O}_{\mathrm{E}}^{-1}\,\{g\} \equiv \frac {1}{\omega_0}\int_{0}^{t}
\sin[\omega_0(t-t')]g(t')dt'.
\end{equation}
Identifying function $g$ as the one given inside the curly braces
in~(\ref{Eq:OE1}), and using $t'= t-\tau$ for convenience, it is easy to obtain
\begin{eqnarray}\label{Eq:xe02}
  x(t) = \hspace{7cm} && \nonumber \\
  -\omega_0 \int_{0}^{t}\sin[\omega_0(\tau)]\sin[\omega_0(t-\tau)]
  \xi(t-\tau)d\tau. \hspace{0.75cm} &&
\end{eqnarray}
We use Eqs.~(\ref{Eq:OE1}) and~(\ref{Eq:OE2}) together with~(\ref{Eq:xe02}) to
get
\begin{eqnarray}\label{Eq:dxdt}
  \bigg\{\frac{d^2}{d t^2}+\omega_0^2\bigg\}x(t) =
  \omega_0 ^{2} \int_{0}^{t}\xi(t)\xi(t-\tau) \hspace{2.cm} && \nonumber \\
  \bigg\{\omega_0 \frac{\sin(2\omega_0\tau)}{2} x(t) -
  \frac{1-\cos(2\omega_0\tau)}{2}\frac{dx(t)}{dt}\bigg\}d\tau, \hspace{0.5cm} &&
\end{eqnarray}
which can be rearranged as
\begin{equation}\label{Eq:AvgRFHO2}
  \frac{d^2}{d t^2}\avg{x} - \omega_0^2 q_1\frac{d}{dt}\avg{x}
  + \omega_0^2 (1 -\omega_0 q_2)\avg{x} = 0,
\end{equation}
where the parameters $q_1$ and $q_2$ are given by\cite{Dykman,Kampen}
\begin{eqnarray}\label{Eq:q1q2}
q_1 &=& 2\int_{0}^{\infty}\avg{\xi(t)\xi(t-\tau)}
[1 -\cos(2 \omega_0 \tau)] d\tau,
\\[0.3cm]
q_2 &=& 2\int_{0}^{\infty}\avg{\xi(t)\xi(t-\tau)} \sin(2 \omega_0\tau) d\tau.
\end{eqnarray}
At this point, we add the term $2\beta d\avg{x}/dt$ to~(\ref{Eq:AvgRFHO2}) and
assuming $\beta \ll \omega_0^{2}$, we may obtain
\begin{eqnarray}\label{Eq:AvgRFHO3}
  \frac{d^2}{d t^2}\avg{x} + (2\beta -
  \omega_0^2 q_1)\frac{d}{dt}\avg{x} && \nonumber \\
  +\,\omega_0^2[1 -\omega_0 q_2]&&\!\!\avg{x} = 0,
\end{eqnarray}
The above equation can also be obtained employing the cumulative expansion of
van Kampen (see Ref.~\refcite{Kampen}).

In the limit of white noise both parameters $q_1$ and $q_2$ vanish. In other
words, white noise does not change the width of the resonance profile of the
frequency. So the rate at which the frequency of the nanocantilever changes due
to mass deposition remains simple to calculate.

With random frequency, the mathematical expression used to find the spectral
densities of noise fluctuations becomes a bit more complicated than that used in
random damping (see Eq.~(\ref{Eq:SDF})). Hence, the analytical
solution~(\ref{Eq:AvgDRHOsol}) for the first moment in random damping can not be
used to obtain a solution for random frequency. However, for dichotomous colored
noise, the amplitude associated with the first moment shows a stochastic
resonance--like feature in $\alpha$  and $\lambda$ similar to that observed by
one of us in an experiment (see Ref.~\refcite{Joshi}). Therefore, a similar
analysis can be applied to get the resonance profile and find the change in
frequency of the nanocantilever due to analyte deposition.

The adsorption-desorption of noise by the nanocantilever gives rise to
fluctuations in its frequency. This is caused mostly by the constant bombardment
of analyte molecules on its surface. This noise mechanism can be understood from
the following perspective. The analyte molecules are adsorbed due to their
affinity to the nanocantilever substrate and are desorbed because of a finite
temperature change. This creates some \emph{fractional} frequency noise. It is
interesting to note that this process of adsorption-desorption of analyte
molecules does not produce a damping mechanism per se. The randomness in the
sticking and releasing particles on the cantilever does not contribute to the
average change in the energy, but it changes the frequency of the nanocantilever
in a nondeterministic manner\cite{Cleland,Ekinci}. Thus, this process introduces
a different parametric noise that does not culminate into dissipation.

The nanoscale cantilevers are quite sensitive to the adsorption-desorption of
noise when compared with conventional scale cantilevers. The reason is the
difference between the surface to volume ratio for each type. This explains why
the number of adsorption locations in nanocantilevers is bigger than those on
the counterpart.

The frequency fluctuations caused by noise can also be described using other
methods different from that used in obtaining~(\ref{Eq:AvgRFHO3}). One of these
methods follows closely Refs.~\refcite{Ekinci} and~\refcite{Cleland}. Let the
adsorption rate be $R_{\mathrm{a}}$, which is dependent on the sticking
coefficient of the nanocantilever surface, and $R_{\mathrm{d}}$ the temperature
dependent desorption rate. The probability of molecular occupation in a
particular area is given by $p=R_{\mathrm{a}}/(R_{\mathrm{a}} +
R_{\mathrm{d}})$, and the corresponding variance in the occupation probability
$\sigma_v=\sqrt{R_{\mathrm{a}} R_{\mathrm{d}}}/(R_{\mathrm{a}} +
R_{\mathrm{d}})$. The correlation time $\tau_{\mathrm{c}}$ of
absorption-desorption noise is also given by $\tau_{\mathrm{c}} =
(R_{\mathrm{a}} + R_{\mathrm{d}})^{-1}$. The spectral density of noise in this
case can be written as
\begin{equation}
  S_{\mathrm{a}}(\omega) = \frac{2 \pi \omega_0^2 N_{\mathrm{a}} \sigma_v^2
  \tau_{\mathrm{c}}}{[1 + (\omega-\omega_0)^2
  \tau_{\mathrm{c}}^2]}\left(\frac{\Delta m}{m}\right)^2,
\end{equation}
where $\Delta m$ is the mass of the molecules attached on the cantilever
surface. The noise variance is a maximum when the probability of occupation is
$1/2$, that is the adsorption rate equals the desorption rate.  On the other
hand, the noise is a minimum when the occupation probability is either $0$ or
$1$. This noise will be superimposed on the frequency change of the biosensor
and hence becoming critical in estimating the analyte mass deposition as we
discuss in the following.

Integration over the spectral density $S_{\mathrm{a}}(\omega)$ provides some
change in the nanocantilever frequency
\begin{eqnarray}
  \Delta \omega_0 &\approx& \left[\int_{\omega_0 - \pi \Delta f} ^{\omega_0 +
  \pi  \Delta f} S_{\mathrm{a}}(\omega)  d\omega \right]^2 \nonumber \\
  &=& - \frac{\omega_0 \sigma_v \Delta m}{2 \pi m}
  [N_{\mathrm{a}} \tan^{-1}(2\pi \Delta f \tau_{\mathrm{c}})]^{1/2},
\end{eqnarray}
where $\Delta f$ defines the width of passband. Hence, the change in mass on the
nanocantilever is given by
\begin{equation}\label{Eq:massDep02}
  \delta m \approx \Delta m \sigma_v
  [N_{\mathrm{a}} \tan^{-1}(2\pi \Delta f \tau_{\mathrm{c}})]^{1/2}.
\end{equation}
Note that this expression for the frequency shift (due to mass deposition) may
be more realistic than that obtained in~(\ref{Eq:massDep01}).

\section{\label{sec:summary}Summary}

In this work, we presented a realistic model for a nanocantilever-based
biosensor using a description of stochastically perturbed harmonic oscillator.
These biosensors work by mass sensing the analytes through shifts in the
characteristic resonance frequency of the oscillator. When analytes are
deposited on the nanocantilever, they bring a variety of noises into the
vibrating system which give rise fluctuations in the damping as well as the
frequency of the cantilever. The estimation of such fluctuations on the spectral
response of the cantilever is important to find out the exact amount of analyte
deposition. This analysis could be used for clinical diagnostic purposes, for
example, early detection of proteins present in cancerous cells. In summary,
this work provides models for damping and frequency fluctuations in a
nanocantilever according to different types of noise and their effect on the
mass deposition.

\nonumsection{Acknowledgments}\noindent
Funding supports from RCSA and URSCA are gratefully acknowledged.


\end{multicols}

\end{document}